\documentclass[useAMS,usenatbib]{mn2e}
\usepackage{graphicx}
\usepackage{natbib}
\usepackage{amsmath}
\usepackage{enumitem}
\usepackage{multirow}
\usepackage{placeins}

\makeatletter
\renewcommand*{\@fnsymbol}[1]{\ensuremath{\ifcase#1\or ^{\dagger}\or ^{*}\or \ddagger\or
   \mathsection\or \mathparagraph\or \|\or **\or \dagger\dagger
   \or \ddagger\ddagger \else\@ctrerr\fi}}
\makeatother

\newcommand{\apj}{ApJ}
\newcommand{\apjl}{ApJL}
\newcommand{\apjs}{ApJS}
\newcommand{\aj}{AJ}
\newcommand{\aap}{A\&A}
\newcommand{\mnras}{MNRAS}
\newcommand{\nat}{Nature}
\newcommand{\na}{New Astron.}

\newcommand{\apss}{Ap\&SS}

\newcommand{\pasp}{PASP}

\newcommand{\pasj}{PASJ}

\newcommand{\aapr}{A\&AR}

\newcommand{\source}{MAXI~J1836$-$194}
\newcommand{\Rin}{$R_{\rm in}$}

\title[\source]{The accretion-ejection coupling in the black hole candidate X-ray binary \source{}}
\author[Russell et al.]{T. D. Russell$^{1}$\thanks{Corresponding e-mail:
thomas.russell@postgrad.curtin.edu.au}, R. Soria$^{1}$, J. C. A. Miller-Jones$^{1}$, P. A. Curran$^{1}$, S. Markoff$^2$, \newauthor D. M. Russell$^{3,4,5}$ and G. R. Sivakoff$^{6}$.\\
$^{1}$International Centre for Radio Astronomy Research - Curtin University, GPO Box U1987, Perth, WA 6845, Australia \\
$^{2}$Astronomical Institute ``Anton Pannekoek'', University of Amsterdam, P.O. Box 94249, 1090 GE Amsterdam, The Netherlands \\
$^{3}$New York University Abu Dhabi, P.O. Box 129188, Abu Dhabi, United Arab Emirates \\
$^{4}$Instituto de Astrof´ısica de Canarias (IAC), E-38200 La Laguna,Tenerife, Spain \\
$^{5}$Departamento de Astrof´ısica, Universidad de La Laguna (ULL), E-38206 La Laguna, Tenerife, Spain \\
$^{6}$Department of Physics, University of Alberta, CCIS 4-181, Edmonton, AB T6G 2E1, Canada \\
} 

\begin{document}

\date{Submitted 2013 November 13}

\pagerange{\pageref{firstpage}--\pageref{lastpage}} \pubyear{2013}

\maketitle

\label{firstpage}

\begin{abstract}

We present the results of our quasi-simultaneous radio, sub-mm, infrared, optical and X-ray study of the Galactic black hole candidate X-ray binary \source{} during its 2011 outburst. We consider the full multi-wavelength spectral evolution of the outburst, investigating whether the evolution of the jet spectral break (the transition between optically-thick and optically-thin synchrotron emission) is caused by any specific properties of the accretion flow. Our observations show that the break does not scale with the X-ray luminosity or with the inner radius of the accretion disk, and is instead likely to be set by much more complex processes. We find that the radius of the acceleration zone at the base of the jet decreases from $\sim 10^6$ gravitational radii during the hard intermediate state to $\sim 10^3$ gravitational radii as the outburst fades (assuming a black hole mass of $8 {\rm M_{\odot}}$), demonstrating that the electrons are accelerated on much larger scales than the radius of the inner accretion disk and that the jet properties change significantly during outburst. From our broadband modelling and high-resolution optical spectra, we argue that early in the outburst, the high-energy synchrotron cooling break was located in the optical band, between $\approx 3.2 \times 10^{14}$ Hz and $4.5 \times 10^{14}$ Hz. We calculate that the jet has a total radiative power of $\approx 3.1 \times 10^{36}$ ergs s$^{-1}$, which is $\sim$6\% of the bolometric radiative luminosity at this time. We discuss how this cooling break may evolve during the outburst, and how that evolution dictates the total jet radiative power. Assuming the source is a stellar-mass black hole with canonical state transitions, from the measured flux and peak temperature of the disk component we constrain the source distance to be 4--10 kpc.

\end{abstract}

\begin{keywords}
accretion, accretion disks -- black hole physics -- stars:individual: MAXI J1836$-$194, outflows -- X-rays: binaries.
\end{keywords}

\section{Introduction}

There is a clear but poorly understood connection between the accretion inflow and jet outflow in accreting compact objects. Low mass X-ray binaries (LMXBs) provide excellent laboratories to probe this relationship as they evolve through their distinct modes of accretion on timescales of weeks and months, allowing us to study the full range of accretion states in the same object on human timescales. Transitions between the canonical states (see \citealt{2005Ap&SS.300..107H} for a full review) are associated with dramatic changes in the structure and power of the outflowing jets \citep{2004MNRAS.355.1105F}. In the hard spectral state, a compact jet is detected and there is a scaling relation between the radio flux and the X-ray luminosity, which, due to an additional mass term, can be extended to active galactic nuclei (AGN) \citep{2003MNRAS.343L..59H,2003MNRAS.345.1057M,2004A&A...414..895F,2012MNRAS.419..267P}. Therefore, studying the inflow-outflow coupling in these objects gives insight into how jets are launched in accreting systems on all physical scales.

While transient black hole (BH) X-ray binaries spend the majority of their lifetimes in a very low-luminosity quiescent state, they occasionally go through an outburst phase where both the radio and X-ray luminosities increase considerably. At the beginning and end of of such an outburst, these systems are observed in a hard state. During the hard state the X-ray spectrum is dominated by a broadband power-law with a weak blackbody component. In the canonical picture the weak blackbody component is thought to originate from a truncated accretion disk and the power law from a geometrically-thick, optically-thin, radiatively-inefficient accretion flow \citep{1996ApJ...465..312E} in the central regions. This state is also associated with a steady, partially self-absorbed, compact jet \citep{2000ApJ...543..373D, 2001MNRAS.327.1273S} with a flat or inverted spectrum \citep{2001MNRAS.322...31F} that is believed to carry away an increasingly dominant fraction of the accretion power as the accretion rate decreases \citep{2003MNRAS.343L..99F}. 

During a typical outburst the system starts in the hard state and, as it brightens, it transits through the hard intermediate state (HIMS) and the soft intermediate state (SIMS), before entering the soft state. During this transition the jet emission evolves into discrete, bright, knots \citep{2004MNRAS.355.1105F} and the compact jet is quenched by at least 2.5 orders of magnitude \citep{2011ApJ...739L..19R,2011MNRAS.414..677C}. In the full soft state most of the accretion power is radiated efficiently by the accretion disk and the X-ray spectrum is dominated by soft, thermal blackbody radiation from a geometrically-thin, optically-thick accretion disk that extends all the way in to the innermost stable circular orbit (ISCO) of the BH \citep{1973A&A....24..337S}.

At the end of the outburst, the X-ray luminosity decreases significantly and the source makes a reverse transition back through the intermediate states into the hard state. During the transition the compact jet is gradually re-established, switching on first in the radio and then in the optical/infrared (IR) bands \citep{2012MNRAS.421..468M}. The gradual jet recovery may be due to an evolution of the jet power over the course of the outburst, as shown by the movement of the jet spectral break, $\nu_{\rm b}$ \citep{2013MNRAS.431L.107C,2013ApJ...768L..35R}, which is the transition between optically-thick and optically-thin synchrotron emission.

While it remains unclear whether the jets are composed of electron/positron pairs or an electron/proton mix \citep[but see][for evidence favouring baryonic jets]{2013Natur.504..260T}, it is clear that electrons are responsible for most of the observed jet radiation through either synchrotron or inverse Compton processes. The flat or inverted radio spectrum (i.e. $\alpha \ga 0$ where $S_{\rm \nu} \propto \nu^{\alpha}$; \citealt{2001MNRAS.322...31F}), which is explained by partially self-absorbed synchrotron emission from electron populations at different distances along the jet, extends up to a frequency above which the jet is no longer self-absorbed. At this frequency the optically thick synchrotron jet spectrum breaks to an optically-thin spectrum, where $\alpha \approx -0.6$ \citep[e.g.][]{2013MNRAS.429..815R}.

The spectral break corresponds to the most compact region in the jet where the electrons are first accelerated from a thermal to a power-law distribution \citep{2001A&A...372L..25M,2005ApJ...635.1203M}, potentially via diffusive shock acceleration \citep[e.g.][]{1978MNRAS.182..147B,1983RPPh...46..973D} though the exact mechanism is not yet known. Typically seen at GHz frequencies in AGN \citep[e.g.][]{1999ApJ...516..672H} and the IR band in hard state LMXBs \citep[e.g.][]{2002ApJ...573L..35C,2011ApJ...740L..13G,2013MNRAS.429..815R,2013ApJ...768L..35R} the frequency at which the break occurs is related to the offset distance from the central BH at which the acceleration occurs and has been found to lie in the range of $10 - 1000 \, r_{\rm g}$ (where $r_{\rm g}$ is the gravitational radius $GM/c^2$, with $G$ as the gravitational constant, $M$ is the BH mass and $c$ is the speed of light) from the central BH for brighter hard states \citep{2001A&A...372L..25M,2003A&A...397..645M,2005ApJ...635.1203M,2007ApJ...670..610M, 2007ApJ...670..600G, 2008ApJ...681..905M, 2009MNRAS.398.1638M}. 

Standard jet theory predicts, in all steady jets from LMXB or AGN, a positive correlation between $\nu_{\rm b}$ and the source luminosity, if the break is associated with the same scale (i.e. distance normalised to $r_{\rm g}$) of the jet at all times \citep{1995A&A...293..665F,2003A&A...397..645M,2003MNRAS.343L..59H,2004A&A...414..895F}. Assuming that a constant fraction of the accretion power is channelled into the jets \citep{1995A&A...293..665F}, the frequency of the spectral break is expected to scale with the mass accretion rate, $\dot{m}$, as $\nu_{\rm b} \propto \dot{m}^{2/3}$. For the radiatively inefficient hard state \citep{2003MNRAS.343L..59H} the X-ray luminosity $L_{\rm X} \propto \dot{m}^2$, thus predicting a relation $\nu_{\rm b} \propto L_{\rm X}^{1/3}$ (see \citealt{2013MNRAS.429..815R} section 3.2, for a full review). 

Recent observations have shown the jet spectral break shifting to lower frequencies during outburst \citep{2013MNRAS.431L.107C,2013ApJ...768L..35R}, with quenching of the radio emission being interpreted as the spectral break moving through the radio band from higher to lower frequencies before returning in the reverse transition.

At higher frequencies (above $\nu_{\rm b}$) a further break can occur in the jet spectrum (to a slope steeper by $\Delta \alpha=0.5$) due to the highest-energy electrons losing a significant fraction of their energy through radiation on timescales that are faster than the dynamical time scale of the source \citep{1998ApJ...497L..17S}. This cooling break ($\nu_{\rm c}$; observed to evolve with time from the X-ray to the optical band in Gamma ray bursts, \citealt{1998ApJ...497L..17S,1998ApJ...500L..97G}), has been suggested to shift from UV energies during low accretion states in LMXBs to X-ray energies at high accretion states \citep{2012ApJ...753..177P,2013MNRAS.429..815R}. However, the evolution of $\nu_{\rm c}$ may also be dependent on other factors, such as the magnetic field \citep[see section 4.2 of][for full discussion]{2013ApJ...773...59P}.

In this paper we present the results of a quasi-simultaneous radio, sub-millimeter, IR, optical, UV and X-ray observing campaign of \source{} during its 2011 outburst. Discovered in outburst on 2011 August 30 with MAXI/GSC \citep{2009PASJ...61..999M} on the International Space Station \citep{2011ATel.3611....1N}, \source{} was classified as a black hole candidate \citep[due to its X-ray and radio properties;][]{2011ATel.3618....1S,2011ATel.3628....1M,2011ATel.3689....1R} with a spin of $a = 0.88\pm0.03$ \citep{2012ApJ...751...34R}. The low-inclination source (between $4^{\circ}$ and $15^{\circ}$, see our companion paper T. D. Russell et al., submitted; hereafter TDR13) underwent a state transition from the hard state to the HIMS on 2011 September 11 \citep[defined by X-ray spectral and timing properties;][]{ferrignoetal2012}. Instead of transitioning through to the full soft state, it reached its softest spectral state on 2011 September 16 where the outburst `failed' \citep{2004NewA....9..249B}; the source hardened, and transitioned back to the canonical hard state on 2011 September 28 before fading towards quiescence \citep{ferrignoetal2012}. \citealt{2013ApJ...768L..35R} (hereafter DMR13) presented the evolution of the compact jet spectrum from radio, IR, optical and UV observations throughout the outburst. Here, we include the X-ray observations and model the full broadband spectra, detailing the changes in the accretion flow that may be driving the evolution of the compact jet.

In Section~\ref{sec:obs} we describe the data collection and reduction. We detail the spectral model we used for our broadband fits in Section~\ref{sec:modelling}. In Section~\ref{sec:results} we present the best fit models for all of our observational epochs and detail the evolution of the source parameters during the outburst and decay, down to an X-ray luminosity an order of magnitude below the peak. In Section~\ref{sec:discussion} we discuss the evolution of the accretion flow and the corresponding changes within the jet that occur during the outburst. We also compare our observations with expected relations and estimate the distance to the source. A summary of results is presented in Section~\ref{sec:summary}.

\section[]{Observations}
\label{sec:obs}

\subsection{VLA}

Following the initial detection of \source{} by MAXI/GSC on 2011 August 30, preliminary radio observations of \source{} were taken with the Karl G. Jansky Very Large Array (VLA) on 2011 September 03 at frequencies of 4.6 and 7.9 GHz. The source was detected at flux densities of $27.5\pm0.6$ and $40.2\pm0.9$ mJy respectively. These detections triggered a campaign of 17 further VLA radio observations under program code 10B$-$218 that continued throughout the outburst and subsequent decay until 2011 December 3. Here we consider only observations that had quasi-simultaneous mid/near-infrared, optical and X-ray data. The simultaneous observations cover the initial rise, transition to the HIMS, return to the hard state and decrease in the radio flux by an order of magnitude.

The radio observations were taken across a range of frequencies from 1 to 43 GHz. The lowest frequency 1--2 GHz data were recorded in two 512 MHz basebands, each of which contained eight 64 MHz sub-bands made up of 64 spectral channels of width 1 MHz. All other frequencies were recorded in 1024 MHz basebands comprised of eight 128 MHz sub-bands, each of which contained 64 spectral channels of width 2 MHz. The data were taken with an integration time of 3 seconds. The array was in the extended A configuration until September 12, after which it was in the A$\to$D move configuration until September 18, and the most compact D configuration thereafter. Due to the presence of a confusing source to the south-west of the target and the compact array configuration used during the bulk of the observations, low-frequency data where the target could not be easily resolved from the confusing source were not considered. 

Data reduction was carried out with the Common Astronomy Software Application ({\small CASA}; \citealt{2007ASPC..376..127M}). Following standard procedures, the data were corrected for shadowing, instrumental issues, and radio frequency interference as well as being Hanning smoothed to minimise the effects of any residual radio frequency interference. Bandpass and amplitude calibration were carried out using 3C286, with the flux scale set using the coefficients derived at the VLA by NRAO staff in 2010 \citep{2013ApJS..204...19P}. Phase calibration was carried out with the nearby compact calibrator source J1820$-$2528. Target observations of between 2 and 4 minutes (depending on the frequency) were sandwiched between two phase calibrator observations, and the amplitude and phase gains derived for the calibrator were interpolated to the target source. Calibrated data were then subjected to multiple rounds of imaging and phase-only self-calibration, with deconvolution being carried out with the multi-frequency synthesis algorithm within {\small CASA}. Flux densities of the source were calculated by fitting a point source to the target in the image plane. At no epoch was the target observed to be significantly extended.

No simultaneous radio data were observed to coincide with the 2011 October 27 mid/near-infrared and X-ray data, but the source was observed on 2011 October 22 and 2011 November 01; we estimate the radio spectrum for this date by fitting the observed exponential decay of the source over the outburst decay for each frequency and interpolating the flux levels for 2011 October 27 (errors were determined from the uncertainty in the exponential fits for each frequency).

\subsection{Submillimetre Array}
\source{} was observed on 2011 September 13 and 2011 September 15 with the Submillimeter Array (SMA). Observations were taken at 256.5 GHz and 266.8 GHz in 2912 MHz basebands. Each of these 2912 MHz basebands were made up of twenty-eight 104 MHz sub-bands. 25 of the 28 sub-bands were comprised of 32 spectral channels of width 3250 kHz, two of the sub-bands were comprised of 512 spectral channels of width 203.125 kHz, and the final sub-band contained 1 channel of width 104 MHz. 

Initial data reduction was carried out with {\small MIRIAD} \citep{1995ASPC...77..433S} to apply a required system temperature correction. The data were then calibrated and imaged following standard {\small CASA} procedures. We used 3C454.3 to calibrate the bandpass, Neptune  to set the flux scale, and the nearby phase calibrators J1911-201 and J1924-292. Target observations of approximately 30 minutes were interleaved between phase calibrator observations. Amplitude and phase calibration were derived for the calibrator sources and interpolated onto the target. \source{} was detected with high significance at a level of $69.7 \pm 6.9$ mJy at 256.5 GHz and $66.4 \pm 6.3$ mJy at 266.8 GHz on 2011 September 13. Phase de-correlation due to poor weather on 2011 September 15 meant that we were unable to place any constraint on the source brightness for this date.  

\subsection{IR/Optical/UV}

Mid-IR observations of \source{} were taken by the Very Large Telescope (VLT) with the VLT Imager and Spectrometer for the mid-Infrared (VISIR) instrument on 4 dates during the 2011 outburst. We observed the system in the {\it PAH1} (8.2-9.0 $\mu$m), {\it SIV} (10.3-10.7 $\mu$m) and {\it J12.2} (11.7-12.2 $\mu$m) filters for all observations, and in the {\it K} (2.0-2.3 $\mu$m) band on some of the dates.

Optical images in the Bessel {\it B, V, R} and Sloan Digital Sky Survey (SDSS) {\it i'} band were taken with the Faulkes Telescopes (North and South) on six dates coincident with the mid-IR and radio data. The UV/Optical Telescope (UVOT) on board the {\it Swift} satellite observed \source{} in six optical and UV filters ({\it v, b, u} and {\it uvw1, uvm2, uvw2}, respectively; \citealt{2008MNRAS.383..627P}) every few days for the duration of the 2011 outburst. {\it Swift}/UVOT data were pre-processed at the Swift Science Data Centre \citep{2010MNRAS.406.1687B} and required only minimum user processing. Spectral files of the {\it Swift}/UVOT data were created with the {\small FTOOL} {\tt uvot2pha}, using an extraction region of 2.5 arcsecond radius.

For full details of the observations and data reduction for all mid-IR to UV observations, see DMR13. 

\subsection{X-ray}

The \textit{Swift} X-ray telescope (XRT) observed \source{} thirty-six times between 2011 August 30 and 2011 November 11, with a total on-source observation time of $\sim 40$ ks. Data were retrieved from the HEASARC public archives. Light curves and spectra, including the background and ancillary response files were extracted with the online XRT data product generator \citep{2009MNRAS.397.1177E}. Suitable spectral response files for single and double events in photon-counting (PC) mode and windowed-timing (WT) mode were downloaded from the latest calibration database. The XRT count rates were high enough ($\rm >1 \; counts \; {s}^{-1}$) to create problems due to pile-up in PC mode, but not in WT mode. Based on our experience with other XRT sources (e.g. \citealt{2011MNRAS.415..410S}), we only fit the WT-mode X-ray data between 0.5 and 10 keV. The {\it Swift}/Burst Alert Telescope ({\it Swift}/BAT) observations were downloaded from the HEASARC public archives and processed with the {\small FTOOL} {\tt batsurvey} to apply standard corrections. The 8-channel spectra and response files were then extracted and {\tt batphasyserr} was used to correct for a standard spectral systematic error. The {\it Swift}/XRT and BAT spectra were both re-binned to a minimum of 20 counts per spectral channel with {\small FTOOLS} so that chi-squared statistics could be used.

The {\it Swift} X-ray light curves were extracted using standard procedures from the XRT online tool\footnote{http://www.swift.ac.uk/user objects/} \citep{2009MNRAS.397.1177E} and the BAT transient monitor\footnote{http://swift.gsfc.nasa.gov/docs/swift/results/transients/} \citep{2013arXiv1309.0755K}.

\section{Spectral Modelling}
\label{sec:modelling}
\label{sec:R_calc}
The complete broadband spectra were modelled with the X-Ray Spectral Fitting Package ({\small XSPEC}; \citealt{1996ASPC..101...17A}) version 12.7. Compatible {\small XSPEC} spectral files of the radio/sub-mm and IR data were created with the {\small FTOOL} {\tt flx2xsp}, allowing for the modelling of the complete broadband spectra, from the radio band through to the X-ray band.

\begin{figure}
\centering
\includegraphics[width=0.42\textwidth]{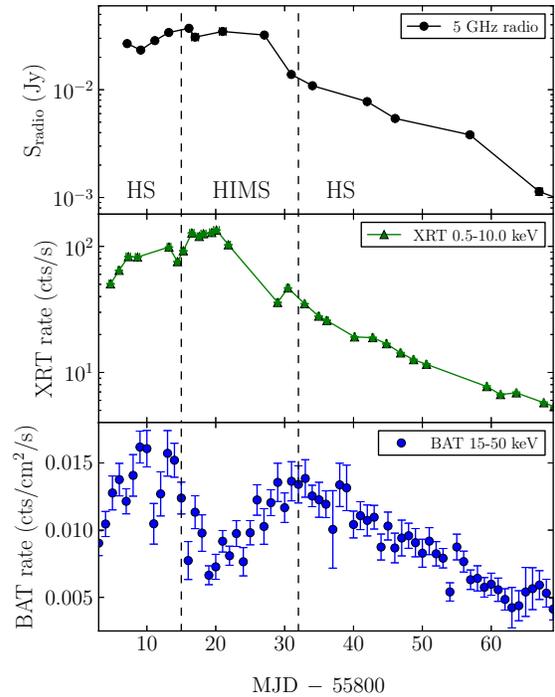}
\caption{Top panel: the 5 GHz radio light curve. Middle panel: the {\it Swift}/XRT (0.5-10.0 keV) light curve. Bottom panel: {\it Swift}/BAT (15-50 keV) light curve. The state transitions are marked by the vertical dashed lines and HS denotes the hard state, while HIMS is the hard intermediate state (states defined by X-ray spectral and timing properties presented by \citealt{ferrignoetal2012}).}
\label{fig:lc_radio_xray}
\end{figure}

\begin{table*}
\caption{Best fitting spectral parameters for the broadband radio to X-ray observations. E(B$-$V)=$0.53_{-0.02}^{+0.03}$ mag was fit as a global parameter. $n_{\rm H}$ refers to the column density along the line of sight. $\alpha_{\rm thick}$ and $\alpha_{\rm thin}$ represent the spectral index of the optically thick and optically thin synchrotron emission, following the convention $\rm S_{\nu} \propto \nu^{\alpha}$. $kT_{\rm disk}$ is the disk temperature in keV. $\Gamma$ is the photon index of the high energy power-law (photon index=1$-$spectral index) simultaneously fitted over the {\it Swift}/XRT and BAT energy range. {\it K} is the normalisation parameter and gives the fitted radius of the inner accretion disk in terms of distance to the source and inclination angle \citep{1998PASJ...50..667K,2007Ap&SS.311..213S}. $y$ gives the log of the outer disk radius in terms of the inner disk radius, i.e. $y \equiv \log(R_{\rm out}/R_{\rm in})$, where $R_{\rm in}$ and $R_{\rm out}$ are calculated from $K$, the inclination angle of the system and the source distance (see Section~\ref{sec:R_calc}). $\nu_{\rm b}$ is the fitted break frequency and $S_{\rm break}$ is the flux density of the break. $\nu_{\rm c}$ is the high energy cooling break of the optically thin synchrotron emission. $R_{\rm in}$ is the calculated inner radius of the accretion disk in $r_{\rm g}$ (assuming a 10 $M_{\odot}$ BH and a distance to source of 8 kpc; see Section~\ref{sec:R_calc} for full details). Errors are 90\% confidence limits and the dates presented are those of the VLA observations.}
\label{tab:xspec_params}
\renewcommand{\arraystretch}{1.3}
\centering
\begin{tabular}{ccccccc}
\hline
Date (VLA) & 2011 Sep 03 & 2011 Sep 12 & 2011 Sep 17 & 2011 Sep 26 & 2011 Oct 12 & 2011 Oct 27 \\
MJD & 55807.12 & 55816.97 & 55821.97 & 55830.95 & 55846.01 & 55861.00 \\

\hline

${n_{\rm H}}$ ($\rm \times 10^{22} cm^{-2}$) & $0.20\pm0.02$ & $0.27\pm0.01$ & $0.29\pm0.01$ & $0.24\pm0.03$ & $0.33\pm0.02$ & $0.39_{-0.02}^{+0.04}$  \\
$\alpha_{\rm thick}$ & $0.70_{-0.09}^{+0.08}$  & $0.20\pm0.02$ & $0.19\pm0.03$ & $0.60_{-0.02}^{+0.05}$ & $0.51_{-0.03}^{+0.04}$ & $0.26\pm0.03$ \\
$\alpha_{\rm thin}$ & $-0.61 \pm 0.1$  & $\le -0.51$ & $-0.70$ -- $-0.33$ & $-0.71_{-0.02}^{+0.03}$ & $-0.76\pm0.03$ & $-0.73_{-0.03}^{+0.06}$ \\
$kT_{\rm disk}$ (keV) & $0.23\pm0.01$  & $0.39\pm0.01$ & $0.42\pm0.01$ & $0.23\pm0.01$ & $0.12_{-0.01}^{+0.02}$ & $0.10\pm0.01$ \\
$\rm \Gamma$ & $1.72 \pm 0.03$  & $1.98\pm0.05$ & $2.03_{-0.06}^{+0.07}$ & $2.01_{-0.06}^{+0.07}$ & $1.77_{-0.04}^{+0.03}$ & $1.78\pm0.05$ \\
{\it K} ($\times 10^{3}$)  & $25.59_{-3.40}^{+3.90}$  & $10.24_{-0.76}^{+0.84}$ & $8.17_{-0.66}^{+0.73}$ & $23.64_{-6.10}^{+9.38}$ & $114.06_{-18.61}^{+31.14}$ & $100.08_{-20.25}^{+35.87}$ \\

{\it y}  & $3.69_{-0.32}^{+0.31}$  & $4.19_{-0.13}^{+0.19}$ & $4.09_{-0.31}^{+0.19}$ & $3.78_{-0.33}^{+0.20}$ & $5.12_{-0.24}^{+0.32}$ & $3.71_{-0.27}^{+0.98}$ \\

$\nu_{\rm b}$ (Hz) & $2.35^{+0.70}_{-1.10}\times 10^{11}$ & $\ge 2.67\times 10^{11}$ & $(0.41$ -- $5.08)\times 10^{11}$ & $9.57^{+8.51}_{-3.62}\times 10^{11}$ & $5.51^{+1.79}_{-0.20}\times 10^{12}$ & $5.09^{+6.80}_{-0.17}\times 10^{13}$ \\

$S_{\rm break}$ (mJy) & $415^{+700}_{-190}$ & $\ge 63$ & 48 -- 80 & $260^{+140}_{-45}$ & $185^{+30}_{-15}$ & $27^{+18}_{-5}$ \\

$\nu_{\rm c}$ (Hz) & $(3.2$ -- $4.5)\times 10^{14}$ & -- & -- & -- & -- & -- \\

$R_{\rm in}$ ($r_{\rm g}$) & $10.2^{+0.9}_{-0.8}$ & $6.5\pm0.3$ & $5.8\pm0.3$ & $9.9^{+1.9}_{-1.3}$ & $21.6^{+3.0}_{-1.8}$ & $20.2^{+3.7}_{-2.1}$ \\

\hline
$\chi^{2}$/d.o.f. & 534.41/443 & 711.59/563 & 565.01/481 & 409.01/484 & 360.95/330  & 287.86/284 \\
\hline
\end{tabular}
\end{table*}

We used a composite of a broken power-law (for the optically thick and thin synchrotron emission) plus irradiated disk model ({\tt diskir}, \citealt{2008MNRAS.388..753G}; for the optical, UV and X-ray bands) to fit the broadband spectra. Absorption in the IR/optical/UV and X-ray bands by gas and dust in the interstellar medium was accounted for with the {\small XSPEC} models {\tt redden} (for the IR/optical/UV band; \citealt{1989ApJ...345..245C}) and {\tt TBabs} (for the X-ray band; \citealt{2000ApJ...542..914W}). 

{\tt diskir} is a Comptonisation model that fits the seed photon spectrum as a standard disk-blackbody. This spectrum is modified by thermal Comptonisation in a hot corona, producing a power-law like component above the peak of the disk emission that is separate from the optically thin synchrotron jet emission. At low energies, the model accounts for the irradiation and re-processing of the X-ray photons in the outer disk and is responsible for the secondary emission bump in the optical/UV band. {\tt diskir} fits the colour temperature $T_{\rm in}$ of the inner disk, the disk-blackbody normalisation constant $K$, the fraction of X-ray flux intercepted and reprocessed in the outer disk, the physical inner radius of the accretion disk ($R_{\rm in}$) and the ratio of outer and inner disk radii ($R_{\rm out}/R_{\rm in}$). An apparent inner disk radius is derived from the normalisation parameter, where $r_{\rm in} \approx D_{\rm 10 kpc} [K/\cos(i)]^{1/2}$ (where $i$ is the inclination angle). The physical inner disk radius is related to the apparent inner radius according to $R_{\rm in} \equiv (\xi^{1/2} \kappa)^2 r_{\rm in} \approx 1.19 r_{\rm in}$ \citep{1995ApJ...445..780S,1998PASJ...50..667K,2007Ap&SS.311..213S}, where $\xi$ is a numerical correction factor to correctly normalise the bolometric disk luminosity, reflecting that $T_{\rm in}$ occurs at a radius larger than $R_{\rm in}$ ($\xi$ is $\sim 0.412$, \citealt{1998PASJ...50..667K}) and $\kappa \sim 1.7$ (e.g. \citealt{1995ApJ...445..780S}) is the ratio of the colour temperature to the effective temperature (the spectral hardening factor). Our composite model does not require any a priori assumptions on whether the near-IR, optical and UV bands are dominated by the optically-thin synchrotron component or by the reprocessed outer disk emission. Magnetic fields are required to launch jets from an accretion flow. The Compton component from {\tt diskir} can, in principle, contain a contribution from thermal electrons, non-thermal electrons, or synchrotron self-Compton due to the presence of a magnetic field. We do not attempt to distinguish between the different Compton components. 

We considered two alternative scenarios for our {\small XSPEC} model; one in which the X-ray power-law component is due to inverse-Compton emission and the other in which it has a significant contribution from the optically-thin synchrotron emission. In the first scenario we truncated the optically-thin synchrotron emission in the far-UV band so that the synchrotron power-law does not contribute significantly to the hard X-ray emission. In the second we left the optically-thin synchrotron power-law unbroken, testing the possibility that the X-ray power-law may be synchrotron rather than Comptonisation (see Section~\ref{sec:source_evolution} for a full discussion). 

\section[]{Results}
\label{sec:results}

\subsection{Light curves and spectral evolution}

\source{} was detected early during its 2011 outburst. During the initial multi-wavelength observation (2011 September 3) the system was observed in the hard X-ray spectral state \citep{ferrignoetal2012}. The system transitioned to the HIMS on 2011 September 11, reaching its softest state on 2011 September 16 (still within the HIMS) before it underwent spectral hardening, transitioning back to the hard state on 2011 September 28 and fading towards quiescence. Similar to other LMXBs in outburst, the source brightened first in the hard X-ray band (15-50 keV) and then in the soft X-ray (0.5-10 keV) and radio bands as it transitioned from the hard state into the HIMS (radio and X-ray light curves presented in Figure~\ref{fig:lc_radio_xray}, full multiwavelength light curves are presented by DMR13, figure 1). At IR, optical and UV wavelengths the source faded during the HIMS, which was attributed to the quenching of the synchrotron component, resulting in a reduction of the emission in those bands (see DMR13 and TDR13, for full description). The outburst did not reach the soft state and the compact jet remained on, hence the continued emission in the radio band.

\begin{figure*}
\centering
\includegraphics[width=0.98\textwidth]{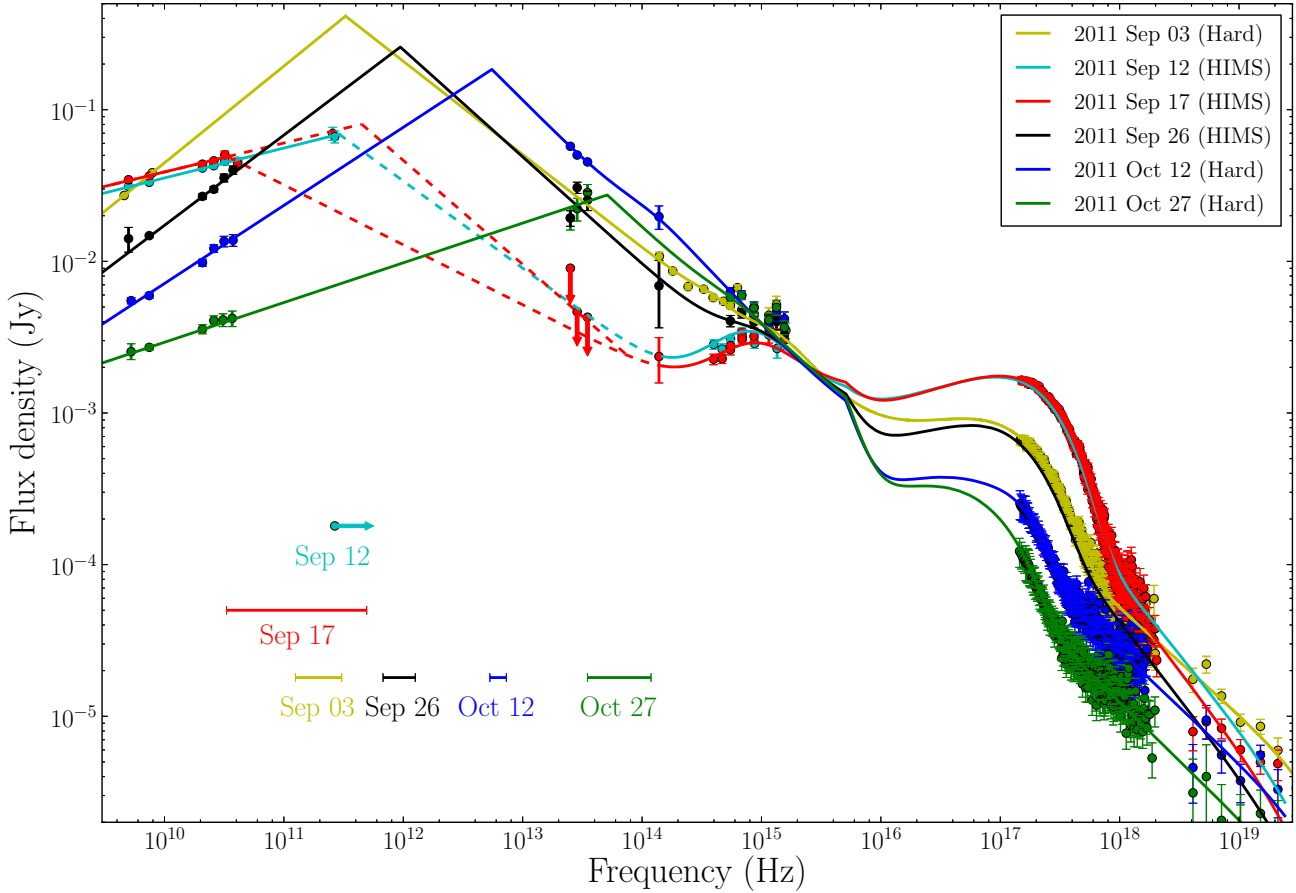}
\caption{Broadband radio to X-ray spectra of \source{} taken during its 2011 outburst. Solid lines represent the XSPEC fits, irradiation of the outer disk causes the slight excess in the IR to UV band. Dashed lines indicate possible ranges of parameters, where the dashed light blue line is the lower limit on the spectral break frequency $\nu_{\rm b}$ for September 12 and the dashed red lines are the lower and upper limits for the September 17 data. Horizontal bars denote the uncertainty range for the spectral break from optically thick to optically thin synchrotron emission. Only four epochs of X-ray observational data (September 3, September 17, October 12 and October 27) are depicted to avoid crowding. The jet spectral break moves to higher (IR) frequencies following the transition back to the hard state during the outburst decay.}
\label{fig:xspec_fits}
\end{figure*}

\subsection{Source evolution}
\label{sec:source_evolution}
While each epoch of the broadband data was fit independently in {\small XSPEC}, absorption in the optical/UV band due to dust in the interstellar medium is not expected to change significantly, therefore the reddening, $E(B-V)$, was fit as a global parameter, tied across all six epochs and found to be $E(B-V)=0.53_{-0.03}^{+0.02}$ mag ($\rm A_V \approx 1.64$ mag). This is in good agreement with estimates from diffuse interstellar bands (from which $E(B-V) \approx 0.6_{-0.1}^{+0.2}$ mag) and results from \citet{2006A&A...453..635M} suggesting there is an absorption layer of $E(B-V)\approx 0.5$ mag in the direction of the source (see TDR13 for further discussion). All other parameters were modelled independently, with the best fitting spectral parameters displayed in Table~\ref{tab:xspec_params} and the broadband models presented in Figure~\ref{fig:xspec_fits}.

\begin{figure}
\centering
\includegraphics[width=0.45\textwidth]{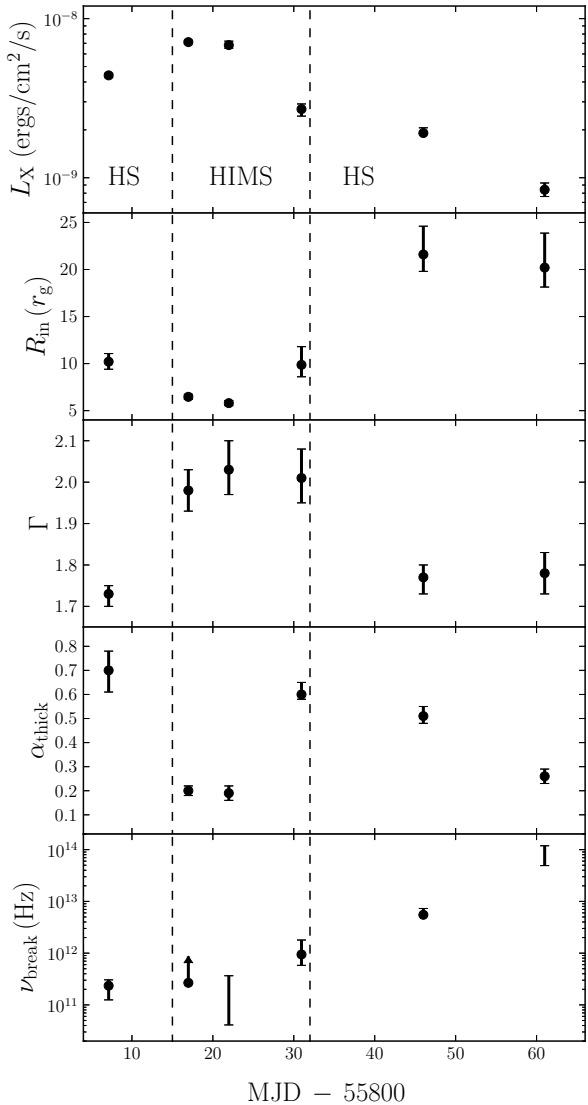}
\caption{Evolution of the system during the outburst, the dashed vertical lines represent the state transitions. Top panel: Source X-ray luminosity (0.5-100.0 keV). Second panel: \Rin{} increases as the system moves from the more disk dominated HIMS to the hard state. Third panel: The X-ray power law component steepened as the disk component increased before decreasing during the decay. Fourth panel: Evolution of the optically-thick spectral index ($\alpha_{\rm thick}$). Bottom panel: The evolution of $\nu_{\rm b}$ during the outburst, shifting by more than two orders of magnitude to higher frequencies as the outburst faded. }
\label{fig:measured_parameters}
\end{figure}

During our first multiwavelength epoch (2011 September 3), taken in the hard state, the X-ray spectrum was dominated by a fairly hard power-law component ($\Gamma = 1.73^{+0.02}_{-0.03}$) combined with a relatively cool accretion disk (Figure~\ref{fig:measured_parameters}) and the radio spectrum was inverted ($\alpha_{\rm thick}=0.70_{-0.09}^{+0.08}$). Our model gives a disk temperature of $0.23 \pm 0.01$ keV and an inner disk radius of $\sim 10 \pm 1$ $r_{\rm g}$ (assuming a $10 \rm M_{\odot}$ BH). During this observation, the break from the optically-thick to optically-thin synchrotron emission was at a relatively low frequency ($\sim 2.3 \times 10^{11} {\rm Hz}$). Our VLT spectra for this epoch (TDR13) show that the synchrotron contribution to the optical/UV continuum must be less than the disk contribution, implying a break in the synchrotron component at lower frequencies. This allows us to constrain the position of the high-energy cooling break to between $3.2 \times 10^{14}$ Hz and $4.5 \times 10^{14}$ Hz (see Section~\ref{sec:highe_break} for further discussion). We were unable to place any constraint on the position of the high energy cooling break for any other epoch and therefore sharply truncate the optically thin synchrotron emission with an exponential cut-off at 20 eV ($\sim 4.84 \times 10^{15}$ Hz, in the far-UV band where we do not have any data available) so that it does not contribute to the X-ray emission. We discuss the alternative possibility that the optically-thin synchrotron emission extends unbroken into the hard X-ray band in Section~\ref{sec:x-ray_synch}.

\begin{table*}
\caption{Best fitting parameters for the broadband observations with an unbroken optically-thin synchrotron power-law. $L_{\rm synch}$/$L_{\rm total}$ is the ratio of the synchrotron emission to the total emission. All other parameters are defined as in Table~\ref{tab:xspec_params}. Errors are 90\% confidence limits and the dates presented are those of the VLA observations.}
\label{tab:xspec_synch_extend}
\renewcommand{\arraystretch}{1.3}
\centering
\begin{tabular}{ccccccc}
\hline
Date (VLA) & 2011 Sep 26 & 2011 Oct 12 & 2011 Oct 27 \\
MJD & 55830.95 & 55846.01 & 55861.00 \\

\hline

$\alpha_{\rm thin}$       & $-0.70_{-0.03}^{+0.02}$ & $-0.76\pm0.03$         & ($-0.73_{-0.03}^{+0.06})^{\footnotesize{a}}$ \\
$kT_{\rm disk}$ (keV)     & $0.24\pm0.02$           & $0.12_{-0.02}^{+0.14}$ & $0.14\pm0.05$\\
$\rm \Gamma$              & $2.3\pm0.2$             & $\le 1.76$             & $(1.94_{-0.25}^{*})^{\footnotesize{b}}$ \\
{\it K} ($\times 10^{3}$) & $19.18_{-6.28}^{+18.20}$& $\le 62.20$            & $100.08_{-20.25}^{+35.87}$ \\
$R_{\rm in}$ ($r_{\rm g}$)& $8.8^{+4.2}_{-1.4}$     & $\le 16.1$             & $20.2^{+3.7}_{-2.1}$ \\
$L_{\rm synch}$/$L_{\rm total}$ (0.5--2 keV) & $\approx 0.15$ & $\approx 0.59$ & $\approx 0.45$  \\
$L_{\rm synch}$/$L_{\rm total}$ (2--10 keV)  & $\approx 0.40$ & $\approx 0.70$ & $\approx 0.84$  \\
$L_{\rm synch}$/$L_{\rm total}$ (10--100 keV)& $\approx 0.72$ & $\approx 0.66$ & $\approx 0.89$ \\

\hline
$\chi^{2}$/d.o.f.         & 403.47/485 & 355.64/330  & 286.91/283 \\
\hline
\multicolumn{4}{l}{\footnotesize{a} The error range for this parameter was constrained from the IR to optical data.} \\
\multicolumn{4}{l}{\footnotesize{b} The asterisk represents where this model is unconstrained.} \\
\end{tabular}
\end{table*}

\source{} was observed three times during the HIMS (on September 12, September 17 and September 26). As the system evolved towards the softest state, the power-law component steepened (reaching a maximum of $\Gamma = 2.03^{+0.07}_{-0.06}$ on September 17) and the disk contribution increased significantly. At this point the disk reached its hottest temperature of 0.42$\pm$0.01 keV, close to the peak colour temperature of $\approx 0.45-0.50 \rm$ keV seen in standard Galactic LMXBs at similar luminosities, and $R_{\rm in}$ reached its minimum value of 5.8 $\pm$0.3 $r_{\rm g}$. During the initial two observations in the HIMS the radio spectrum flattened (to $\alpha_{\rm thick}\sim 0.2$, see Table~\ref{tab:xspec_params} and Figure~\ref{fig:xspec_fits}) before it became more inverted prior to the transition back to the hard state ($\alpha_{\rm thick}$ was observed to be $0.6^{+0.05}_{-0.02}$ on September 26). Due to sparse IR data during the September 12 and September 17 observations we can only place limits on the position of the break frequency. For September 12 we determined that the break frequency must lie above the sub-mm detection, at $\ge 2.67\times 10^{11}$ Hz, and for September 17 the jet break occurred between 0.41$\times 10^{11}$ and 5.08$\times 10^{11}$ Hz (with the lower limit set by requiring the break to lie above the maximum observed radio frequency, and the upper limit dictated by the K-band detection and the lower frequency IR band upper limits, allowing a minimum optically-thin slope of $\alpha > -0.70$). 

After September 17, the outburst faded as the source underwent spectral hardening. \source{} transitioned back to the hard state on September 28, following which the power-law component hardened (to $\Gamma = 1.78 \pm 0.05$ on October 27) and the disk contribution decreased, consistent with a typical transition from the HIMS to the hard state (the disk temperature decreased to $0.10 \pm 0.01$ keV and the inner disk radius increased to $R_{\rm in}=20^{+4}_{-2} \, r_{\rm g}$ on October 27). Following the transition back to the hard state the spectral index of the optically-thick synchrotron spectrum remained fairly constant, consistent with an inverted spectrum. As the system settled in the canonical hard state we observed the jet spectral break shift to much higher frequencies (over two orders of magnitude, to $\sim 5 \times 10^{13}$ Hz by October 27) as the radio spectrum flattened to $\alpha_{\rm thick} = 0.26\pm0.03$. 

Our best fitting spectral parameters are typical of LMXBs transitioning between the hard state and the HIMS \citep[e.g.][]{2005Ap&SS.300..107H} and are in good agreement with those presented by \citet{ferrignoetal2012}. Any discrepancy between them is a result of differences between the data modelled, as well as differences between the physical Comptonisation model and a phenomenological model where the X-ray power-law is not truncated at low energies.

\subsubsection{X-ray synchrotron model}
\label{sec:x-ray_synch}
For the epochs where we do not have any constraint on the position of the cooling break, we considered the possibility that the synchrotron power-law component may have extended unbroken into the hard X-ray band. Due to poor IR data constraints during our September 12 and 17 observations there is a large possible range of slopes for the optically-thin synchrotron power law. As a result the synchrotron component will either not contribute significantly to the X-ray emission (dominated by a Comptonised disk), or may require a break in the X-ray band due to the observed 2--10~keV photon index being steeper than the optically-thin synchrotron slope. Therefore, we did not apply this model to these two epochs.
  
For the final three observational epochs (September 26, October 12 and October 27) we fit the data without truncating the optically-thin synchrotron emission in the UV-band (Table~\ref{tab:xspec_synch_extend}). For September 26, we cannot rule out the synchrotron power-law extending unbroken into the X-ray band. In that case we find that the synchrotron component can account for $\approx 15\%$ of the 0.5--2~keV emission, $\approx 40\%$ of the 2--10~keV emission and $\approx 72\%$ of the 10--100~keV emission (the rest being due to Comptonisation). For the October 12 and 27 epochs, the slope and normalisation of the optically-thin synchrotron emission and the high-energy (hard) X-ray component are very similar. The unbroken synchrotron component accounts for the majority of the emission above 2 keV (Table~\ref{tab:xspec_synch_extend}).

We find that both models (with and without a cooling break) are statistically equivalent (from an F-test). Therefore, we cannot rule out either model with the data on hand. We therefore adopt the conservative scenario in which the hard X-ray emission comes from inverse Compton in a hot corona \citep[e.g.][]{2007A&ARv..15....1D,2009MNRAS.400.1512M,2012IJMPS...8...73M}. A full investigation of the alternative scenario (where the X-ray emission at energies as high as 100 keV comes from synchrotron emission; e.g. \citealt{2010MNRAS.405.1759R}) is beyond the scope of this work.

\section{Discussion}
\label{sec:discussion}
We have observed the gradual evolution of the jet and the accretion disk in \source{} during its 2011 outburst. As reported by DMR13, the spectral break between the optically thick and optically thin synchrotron emission evolved to higher frequencies as the jet recovered and the accretion disk faded following the outburst. In the following section we discuss how the accretion flow may be coupled to the compact jet and compare our observations with expected relations.

\subsection{Break frequency}
\label{sec:nu_b}
A shifting jet spectral break has now been observed during single outbursts of two LMXBs; this source (DMR13) and GX 339-4 \citep{2011ApJ...740L..13G,2013MNRAS.431L.107C}. DMR13 presented radio to optical observations of \source{} during outburst, showing that the break moved to higher frequencies by more than two orders of magnitude as the outburst faded. The radio spectrum of GX 339$-$4 switched from optically-thin to optically-thick synchrotron emission during the decay phase of its 2010$-$2011 outburst \citep{2013MNRAS.431L.107C}, implying that the jet spectral break transitioned through the radio band to higher frequencies as the outburst faded. As a typical LMXB outburst fades, the IR flux is initially seen to increase (e.g. \citealt{2005ApJ...622..508K,2007MNRAS.379.1401R,2009MNRAS.400..123C,2012AJ....143..130B,2013MNRAS.431L.107C}) as $\nu_{\rm b}$ shifts to higher frequencies. This suggests that the movement of $\nu_{\rm b}$ to lower/higher frequency and the IR fading/brightening during an outburst is direct evidence for the quenching and recovery of the compact jet, but the accretion processes that may be responsible for this evolution are yet to be determined.    

\begin{figure}
\centering
\includegraphics[width=0.42\textwidth]{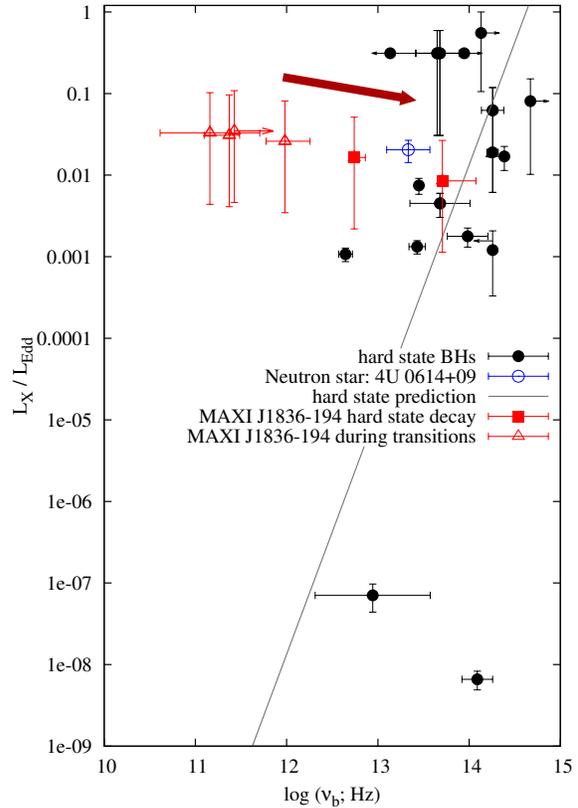}
\caption{Jet break frequency versus X-ray luminosity as a fraction of the Eddington luminosity for hard state and quiescent black hole X-ray binaries \citep{2013MNRAS.429..815R} and the neutron star 4U 0614$+$09 \citep{2010ApJ...710..117M}, with the theoretically expected $\nu_{\rm b} \propto L_{\rm X}^{1/3}$ relation (grey line). Results from the 2011 outburst of \source{} have been included, where the solid red squares show the position during the hard state and the red triangles for the HIMS and the transitioning September 03 epoch (assuming a distance to the source of 8 kpc and a BH mass of 10 M$_{\odot}$). This shows that the $\nu_{\rm b}$ in \source{} shifted to much lower frequencies during the HIMS and transitioned back as the outburst faded (the evolution of $\nu_{\rm b}$ is outlined by the arrow).}
\label{fig:canonical_jet_break}
\end{figure}

Theoretical relations predict that during the hard state the break frequency should scale with the X-ray luminosity as $\nu_{\rm b} \propto L_{\rm X}^{1/3}$, derived from the mass accretion rate of a radiatively inefficient flow (see \citealt{2013MNRAS.429..815R}, section 3.2.1 for a full review). We observe a clear discrepancy with that relation following the transition back to the hard state; during the decay phase $\nu_{\rm break}$ increased by more than two orders of magnitude as $L_{\rm X}$ decreased (Figure~\ref{fig:measured_parameters}). In Figure~\ref{fig:canonical_jet_break}, we plot the break frequency against X-ray luminosity for \source{} as well as a number of other LMXBs \citep{2013MNRAS.429..815R}, illustrating the motion almost perpendicular to the expected $\nu_{\rm b} \propto L_{\rm X}^{1/3}$ relationship. The observed scatter of the jet break frequencies in different sources at similar luminosities rules out a direct global relation between different LMXBs \citep{2013MNRAS.429..815R}. This means that other accretion parameters (that are different between sources) such as inner disk radius, disk temperature, magnetic field strength and BH mass may have a significant effect on $\nu_{\rm b}$. We show that for an individual source (i.e. at constant mass) during spectral state transitions the jet break frequency is being driven primarily by the changing structure of the accretion flow, rather than by luminosity. The spectral break from our final epoch is seen at frequencies comparable to other LMXBs and we speculate that the expected relation may hold once the system has fully settled into a canonical hard state. If the jet has settled into a canonical hard state by our final observational epoch, we can place a minimum timescale of $\approx$~29~days for the jet to settle into the hard state following the X-ray defined state transition.

\begin{figure}
\centering
\includegraphics[width=0.45\textwidth]{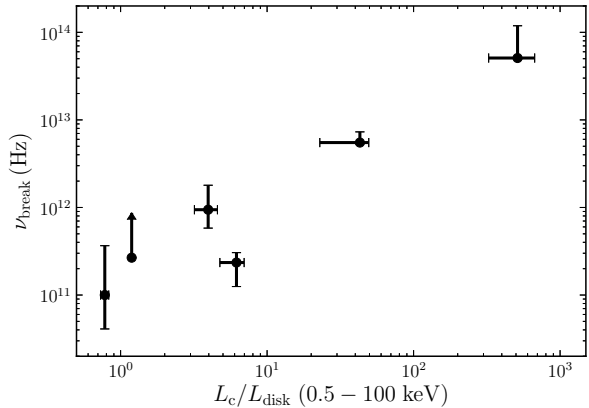}
\caption{Evolution of the spectral break frequency with spectral hardness, parameterised by the ratio of Comptonised flux to disk flux, $L_{\rm c}/L_{\rm disk}$, over the 0.5--100 keV band. Performing a Monte Carlo Spearman's rank correlation test produces a Spearman's rank correlation coefficient of $0.71\pm0.14$ at a significance level of $0.86\pm0.12$. }
\label{fig:hardness_correlation}
\end{figure}

We have shown here that the position of the jet break is not driven by the X-ray luminosity alone, but must also be influenced by other accretion parameters (such as $R_{\rm in}$ and spectral hardness). While the transition to higher frequencies did occur as the inner radius of the accretion disk shifted outwards there does not appear to be a strong relation between $\nu_{\rm b}$ and $R_{\rm in}$ (Figure~\ref{fig:measured_parameters}). However, the position of the jet break does appear, in part, to be related to the conditions in the X-ray emitting plasma that also drive the source hardness. Performing a Monte Carlo Spearman's rank correlation of the jet break frequency against hardness provides tentative evidence for a positive correlation between the two (Figure~\ref{fig:hardness_correlation}), with a Spearman's rank coefficient of $0.71\pm0.14$ at a significance level of $0.86\pm0.12$. For this test hardness was defined as the ratio of Comptonised flux ($L_{\rm c}$) to disk flux ($L_{\rm disk}$) across the 0.5--100 keV band. Recent results, presented by \citet{2013arXiv1308.4332V}, showed that MAXI J1659$-$152 reached a full soft state and its $\nu_{\rm b}$ shifted to lower frequencies than we observe here, further suggesting that the source hardness may be correlated with the break frequency in some way. We also observed $\nu_{\rm b}$ shift to lower frequencies as the X-ray photon index steepened; however the change in the photon index component only occurred over a very small range (values range from $\Gamma \approx 1.73$ to $\approx2.03$, see Figure~\ref{fig:measured_parameters}) and there was no significant correlation between $\Gamma$ and $\nu_{\rm b}$ (a Monte Carlo Spearman's rank correlation produces a correlation coefficient of $-0.26\pm0.20$ at a significance level of $0.40\pm0.25$).

While the correlation test provides some evidence for source hardness being correlated with $\nu_{\rm b}$, the physical mechanisms that set the position of the first acceleration zone (that is associated with $\nu_{\rm b}$) are likely to involve more complex processes \citep[e.g. set by MHD and plasma processes;][Polko et al., in press]{2010ApJ...723.1343P,2013MNRAS.428..587P}.

\subsection{First acceleration zone}
\label{sec:faz}

It is thought that the compact jet is launched on scales of a few to 100 $r_{\rm g}$ (comparable to $R_{\rm in}$; \citealt{2005ApJ...635.1203M, 2012ApJ...753..177P}). Following the standard single-zone synchrotron theory presented by \citet{2011A&A...529A...3C}, we calculate the evolution of the radius of the region where electrons are first accelerated up into a power law distribution ($R_{\rm F}$, which is defined by the cross-sectional area of this region), as well as an estimate of the equipartition magnetic field ($B_{\rm F}$).
 
\begin{figure}
\centering
\includegraphics[width=0.45\textwidth]{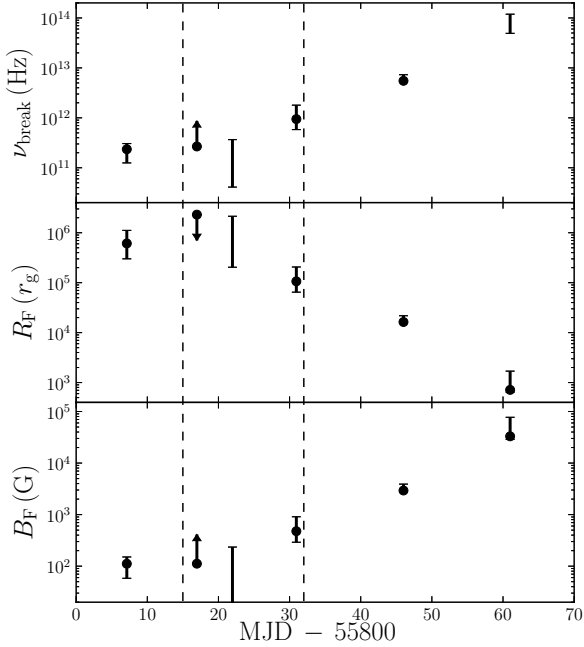}
\caption{Evolution of the jet acceleration region, the dashed vertical lines represent the state transitions. Top panel: The evolution of $\nu_{\rm b}$ during the outburst. Second panel: Radius of the first acceleration zone, $R_{\rm F}$ (in $r_{\rm g}$ and assuming a $10 \rm M_\odot$ BH), reducing as the source decays. Bottom panel: The magnetic field at the jet base ($B_{\rm F}$) increases as the source transitions from the HIMS to the hard state and decays. }
\label{fig:calculated_parameters}
\end{figure}

$R_{\rm F}$ is determined by the position and frequency of the spectral break, as well as the spectral index of the optically thin synchrotron emission, according to
\begin{equation}
R_{\rm F} \propto \nu_{\rm b}^{-1} S_{\nu_{\rm b}}^{(p+6)/(2p+13)},
\label{eq:R_F}
\end{equation} 
where {\it p} is the slope of the electron energy spectrum ($p=1-2\alpha$, following the convention $S_{\nu} \propto \nu^{\alpha}$) and $S_{\nu_{\rm b}}$ is the flux density at the jet spectral break. Evaluating this expression for each epoch at which we constrain the break frequency, we find that $R_{\rm F}$ was largest during the HIMS ($\sim10^5-10^6 r_{\rm g}$) and moved inwards (to $\sim 10^2 - 10^3 r_{\rm g}$) as the outburst faded in the hard state (Figure~\ref{fig:calculated_parameters}). Our results imply that the particle acceleration occurs at much larger scales than $R_{\rm in}$ and that the jet properties are changing significantly during the outburst. We infer that during outburst the jet luminosity is reduced as the particle acceleration occurs at much larger radii, and then recovers as the acceleration point shifts inwards (demonstrated by $\nu_{\rm b}$ moving to higher frequencies) and reduces in size (demonstrated by the reduction of $R_{\rm F}$). While the particle acceleration is occurring on different scales in the jet, we are unable to say if the jet is thermally dominated at smaller scales. While this model does demonstrate the evolution of the accelerating region, more detailed treatments (e.g. \citealt{2001A&A...372L..25M,2010ApJ...723.1343P,2013MNRAS.428..587P}) are required (although the final results may only be expected to differ by a factor of a few). 

The magnetic field in the accelerating region is expected to increase as $R_{\rm F}$ moves inwards. This is due to the magnetic field and density increasing as we probe more compact regions of the jet. An estimate of the equipartition magnetic field can also be calculated from the observed jet break, according to
\begin{equation}
B_{\rm F} \propto \nu_{\rm b} S_{\nu_{\rm b}}^{-2/(2p+13)}.
\label{eq:B_F}
\end{equation} 
We find that the magnetic field in the acceleration region increased from $\sim 10^2$ G when $R_{\rm F}$ was large to $\sim 5 \times 10^4$ G  when the acceleration occurred closer in to the BH (Figure~\ref{fig:calculated_parameters}). 

\subsection{Radio spectrum}

The spectral index of the optically-thick synchrotron emission is thought to depend on the geometry of the compact jet, the bulk flow velocity as a function of distance along the jet \citep{1979ApJ...232...34B,1995A&A...293..665F,2006MNRAS.367.1083K,2009ApJ...699.1919P}, and the injection of energy from, e.g., shocks and turbulence \citep{2006MNRAS.367.1083K,2010MNRAS.401..394J}. 

Our observations show that the radio spectrum, which was inverted in our initial observation, flattened during the HIMS before becoming more inverted prior to the transition back to the hard state. Following that state transition $\alpha_{\rm thick}$ remained steady for a few weeks until flattening once again (Figure~\ref{fig:measured_parameters}, fourth panel). Although $\alpha_{\rm thick}$ flattened during the HIMS, there is no evidence of a significant correlation with any of the other measured system parameters. While our current data do not allow us to determine which parameters are driving the evolution of the radio spectrum, it is clear that the jet properties are changing significantly during the outburst.

\subsection{High-energy cooling break}

\label{sec:highe_break}
\begin{figure}
\centering
\includegraphics[width=0.45\textwidth]{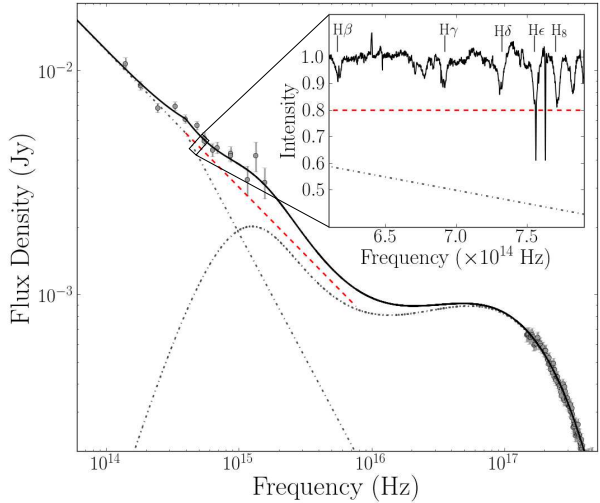}
\caption{Main axes: Unfolded best fit {\small XSPEC} model of the September 3 data, showing the contribution of the optically-thin synchrotron emission with the irradiated disk. The high-energy cooling break lies at $\approx 3.2-4.5\times10^{14}$ Hz. The dash-dotted grey lines represent the separate jet and irradiated disk components, the (red) dashed line represents the extrapolated optically thin power law and the solid line is the total observed spectrum. Inset: The averaged normalised blue spectrum from August 31 and September 1 with a selection of the Balmer lines labelled (TDR13). The red dashed line is the extrapolated optically thin power law (seen in the main axes), accounting for $\sim 80\%$ of the optical continuum (a similar level as the Balmer absorption lines). The dash-dotted grey line is the optically thin synchrotron emission after the placement of the cooling break. The NIR, optical, UV and X-ray data points are shown as the light grey points. This demonstrates that the cooling break must occur below this frequency band if the Balmer absorption lines are not to span the entire disk continuum.}
\label{fig:high_energy_break}
\end{figure}

If we simply extrapolate the optically-thin power-law from the IR into the optical band, the synchrotron component would account for at least 80\% of the optical continuum during our September 3 epoch. However, the VLT spectra taken on 2011 August 31 and September 1 (presented in TDR13 and taken within a day of the NIR and optical data for our September 3 epoch) show prominent higher-order Balmer absorption lines ($\rm H_{\gamma}$, $\rm H_{\delta}$, $\rm H_{\epsilon}$, $\rm H_{8}$ and $\rm H_{9}$) with a relative depth up to 20\% of the continuum flux, which is typical of optically-thick accretion-disk spectra (see inset in Figure~\ref{fig:high_energy_break}); similar absorption lines are often seen for example in Cataclysmic Variable disks at the onset of an outburst, and particularly in low-inclination systems \citep{2003cvs..book.....W}. Strong Balmer absorption lines are inconsistent with a continuum dominated by synchrotron emission (the Balmer lines, which increase in depth at shorter wavelengths, would then span the full depth of the disk continuum; \citealt{1980MNRAS.193..793M,1993PASP..105..761W}). The only way to reconcile the synchrotron power-law spectrum in the near IR with a disk component large enough to accommodate the depths of the Balmer lines in the blue optical region is if the expected high-energy spectral break of the synchrotron emission is located in the red part of the optical band. We tested this by allowing the cooling break to be at lower frequencies and refit the data.

Specifically, we find good fits when the break is located between $\approx 3.2 \times 10^{14}$ Hz and $\approx 4.5 \times 10^{14}$ Hz, meaning that at this point in the outburst the disk is contributing around 50\% of the optical emission (with the depth of the H$_{\epsilon}$ and $\rm H_{8}$ Balmer absorption lines now $\sim$40\% of the disk continuum, consistent with low-inclination accretion disks; \citealt{1980MNRAS.193..793M}).

The high energy synchrotron break, $\nu_{\rm c}$, results from the rapid radiative cooling (faster than the dynamical timescale) of the electrons when they are no longer being continuously accelerated. The position and evolution of $\nu_{\rm c}$ dictates the total jet radiative power and is important to understanding the outflows from LMXBs. For example, it is still debated whether the high energy (hard) X-ray emission in the low/hard state of LMXBs is dominated by the emission from the outer jet, by synchrotron self-Compton from the base of the jet/corona, or by the inverse-Compton scattering of seed disk photons. Recent work discussing the evolution of the X-ray (and jet synchrotron) spectral index (see \citealt{2012ApJ...753..177P,2013MNRAS.429..815R,2013MNRAS.434.2696S}) suggests that $\nu_{\rm c}$ may occur at high X-ray energies ($\ga 10$keV) at the high-luminosity end of the low/hard state following an outburst ($\sim 10^{-3} L_{\rm Edd}$), before shifting to the UV band as the system moves into quiescence ($\sim 10^{-5} L_{\rm Edd}$), implying that the jet may dominate the X-ray emission in quiescence \citep{2013ApJ...773...59P}. However, it has also been argued that the observed change in X-ray spectral index can also be quantitatively explained by a truncated disc with a radiatively inefficient hot inner flow \citep{2013MNRAS.434.3454G}, implying a cooling break below X-ray energies.

Our observations require the synchrotron cooling break to lie at even lower frequencies, in the optical band, in the first few days of the outburst (when the bolometric luminosity is a few $\times 10^{-2} L_{\rm Edd}$). This suggests that the jet is already evolving early in the outburst, at the same time as the system is brightening and the disk begins filling in. The apparent similar evolution of $\nu_{\rm b}$ and $\nu_{\rm c}$ to lower frequencies (assuming that $\nu_{\rm c}$ occurs in the X-ray band in the low/hard state; \citealt{2013MNRAS.429..815R}) would suggest that the evolution of $\nu_{\rm c}$ may be coupled with $\nu_{\rm b}$. As $\nu_{\rm b}$ shifts to lower frequencies during spectral softening the electrons are accelerated from much further out (which we discuss in Section~\ref{sec:faz}) and therefore cannot be accelerated up to X-ray energies. Unfortunately we are unable to place any constraint on $\nu_{\rm c}$ (except that $\nu_{\rm c} > \nu_{\rm b}$) at any other time and therefore cannot speculate further on the evolution as the accretion rate changes.

During our final two observational epochs (October 12 and 27; during the decay phase of the outburst), the measured high-energy (hard) X-ray spectral slope is observed to be very similar to the slope of the optically thin synchrotron emission, and even its normalisation is consistent with an extrapolation of the radio/IR power-law to the X-ray band (in fact, for these two epochs the hard X-ray emission can also be plausibly fit with an extension of the synchrotron power-law; see Section~\ref{sec:x-ray_synch}). This would be an intriguing coincidence if the two components had different origins (synchrotron and inverse Compton). Thus, although this does not prove a common origin, it is plausible that the hard X-ray emission is dominated by synchrotron emission from the jet when the system re-enters the low/hard state. A similar result was also observed in the outburst decay of XTE~J1550--564, where the slope of the optically-thin synchrotron emission was found to be consistent with that of the hard X-ray power-law, suggesting a single unbroken optically-thin power-law extending from the IR band to the hard X-ray band \citep{2010MNRAS.405.1759R}. Extrapolating the optically-thin power-law to X-ray energies also agrees with the observed X-ray luminosity in a number of other BH LMXBs at similar Eddington luminosities in the hard state \citep[see figure 5 of][]{2013MNRAS.429..815R}. Here, we speculate that $\nu_{\rm c}$ shifts again across the UV band towards the X-ray band as the outburst decays further as the disk cools down or disappears and continuous jet acceleration is re-established.

\subsection{Radiative jet luminosity}

With knowledge of the cooling break frequency, the total radiative jet luminosity for our initial epoch (September 3) can be calculated from the complete jet spectrum. By integrating across the entire jet spectrum (up to infinity)  we find $L_{\rm Jet, rad} \approx 1.8\times 10^{36}$ erg s$^{-1}$, assuming a source distance, $d$, of 8 kpc (for further discussion on assumed distance see Section~\ref{sec:distance}). Using a canonical radiative to kinetic energy ratio of $\approx 5\%$ \citep{1979ApJ...232...34B,2000MNRAS.318L...1F,2001MNRAS.322...31F,2005Natur.436..819G,2007MNRAS.376.1341R} the total inferred jet power $P_{\rm Jet} \approx 3.6 \times 10^{37}$ erg s$^{-1}$~($d$/8 kpc)$^2$. From the broadband {\small XSPEC} models, the bolometric luminosity at that epoch was $L_{\rm Bol}\approx 4.9 \times 10^{37}$~erg~s$^{-1}$ ($2\times 10^{-8}$~--~200 keV), indicating that the jet is $\sim$ 3 \% of $L_{\rm Bol}$ if purely radiative. However, the jet is oriented close to the line of sight (TDR13), meaning it will be significantly Doppler boosted. $S = S_{\rm 0} \delta^{3-\alpha}$ (where $\delta$ is the relativistic Doppler shift), so for a canonical bulk Lorentz factor of 2, the intrinsic jet power is reduced (by a factor of $\sim 12$) to $P_{\rm Jet}\approx3.1\times10^{36}$ erg s$^{-1}$ ($d$/8 kpc)$^2$. In agreement with findings presented by \citet{2003MNRAS.343L..99F}, this result demonstrates that the system is not jet-dominated at high luminosities.

\begin{figure}
\centering
\includegraphics[width=0.43\textwidth]{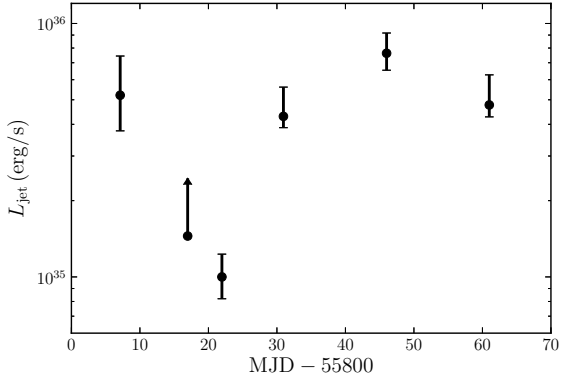}
\caption{The fading and recovery of the compact jet, assuming a distance to source of 8 kpc. Minimum estimate for the total radiative jet luminosity calculated by integrating across the observed jet band ($5 \times 10^9$ Hz to $7 \times 10^{14}$ Hz). The minimum radiative luminosity decreases as the system reaches its softest state, and then recovers on spectral hardening.}
\label{fig:jet_luminosity}
\end{figure}

For all other epochs we are unable to place any observational constraints on $\nu_{\rm c}$ and therefore can only place a lower limit on the jet luminosity. Integrating across the observable jet spectrum (up to $7 \times 10^{14}$ Hz), we find that the minimum radiative luminosity of the compact jet marginally increases as the system spectrally hardens as it transitions from the HIMS to the hard state (Figure~\ref{fig:jet_luminosity}; see DMR13 for full analysis). The true radiative jet luminosity will depend on the position of $\nu_{\rm c}$, which will significantly overshadow the observed quenching and recovery of the minimum radiative jet power in Figure~\ref{fig:jet_luminosity} (shown by the factor of $\sim$6 difference between the total jet power and minimum radiative luminosity for September 3). 

Assuming that $\nu_{\rm c}$ shifts through the X-ray band by October 27 and lies above XRT energies (at $\sim 2 \times 10^{18}$), then $L_{\rm Jet,rad} \approx 6 \times 10^{36}$ erg s$^{-1}$ (d/8 kpc)$^2$. The bolometric luminosity for Oct 27 is $L_{\rm Bol}\approx 1.1 \times 10^{37}$ erg s$^{-1}$ ($2\times 10^{-8}$--200 keV). Therefore, the jet accounts for $\sim$ 60 \% of $L_{\rm Bol}$ if purely radiative. Assuming the 5\% radiative to kinetic ratio and accounting for Doppler boosting, the total power of the jet is $P_{\rm Jet}\approx 1 \times 10^{37}$ erg s$^{-1}$ (d/8 kpc)$^2$, which is a factor of $\sim$20 higher than our minimum radiative energy calculation (Figure~\ref{fig:jet_luminosity}). This result demonstrates that the system would become jet dominated should $\nu_{\rm c}$ shift to higher frequencies.

\subsection{Distance to the source}
\label{sec:distance}
From the measured values of inner-disk temperature and the X-ray state evolution we can estimate plausible values of source distance for a range of BH masses. In a typical outburst only minor spectral softening occurs beyond the HIMS \citep{2010LNP...794.....B} and there is evidence that, even in the low/hard state, at luminosities $\sim$0.01 $L_{\rm Edd}$ the inner accretion disk extends close to the ISCO \citep[e.g.][]{2006ApJ...653..525M,2007Ap&SS.311..149M,2009MNRAS.397..666W,2010MNRAS.402..836R,2010ApJ...716.1431R,2011MNRAS.414L..60U,2012ApJ...757...11M, 2013ApJ...769...16R}. \citet{2011MNRAS.415..410S} also observed the inner-disk temperature to remain approximately constant during transitions between the intermediate states and soft states of GRS 1758$-$258. Therefore, we assume that during our softest spectral observation $R_{\rm in} \approx R_{\rm ISCO}$. At this point in time the disk reached a temperature of $0.42 \pm 0.01$ keV, close to the peak colour temperature of $\approx 0.45-0.50$ keV seen in standard Galactic LMXBs at comparable values of $L_{\rm X}/L_{\rm Edd}$. 

For a standard Shakura-Sunyaev disk, the bolometric disk luminosity is related to the inner disk temperature and the apparent inner disk radius by 
\begin{equation}
L_{\rm disk} = 4 \pi \sigma T_{\rm in}^4 r_{\rm in}^2. 
\label{eq:Ldisk_r}
\end{equation}

Following \citet{2007Ap&SS.311..213S}, as $R_{\rm in} \approx R_{\rm ISCO} \approx 6 \phi GM/c^2$, where $\phi$ depends on the BH spin ($\phi=1$ for a Schwarzschild BH and $\phi=1/6$ for an extreme Kerr BH) and from the $R_{\rm in}$ to $r_{\rm in}$ relation (see Section~\ref{sec:R_calc}) we find that the mass of the BH can be approximated by
\begin{multline}
M \approx 10 \left( \frac{\eta}{0.1} \right) \left( \frac{\xi \kappa^2}{1.19} \right) \left( \frac{L_{\rm disk}}{5 \times 10^{38} {\rm erg \, s^{-1}}} \right)^{1/2} \\
\times \left( \frac{kT_{\rm in}}{1 {\rm keV}} \right)^{-2} M_{\odot} \\
\label{eq:M_Ldisk}
\end{multline}
where $\sigma$ is the Stefan-Boltzmann constant, $\eta$ is the radiative efficiency $\approx 1/12\phi$ in a Newtonian approximation, $\xi=0.412$ which reflects that $T_{\rm max}$ occurs at $R=(49/36) R_{\rm in}$ \citep{1998PASJ...50..667K,2002apa..book.....F} and $\kappa \sim 1.7$ (e.g. \citealt{1995ApJ...445..780S}) is the ratio of the colour temperature to the effective temperature (the spectral hardening factor).

\begin{figure}
\centering
\includegraphics[width=0.43\textwidth]{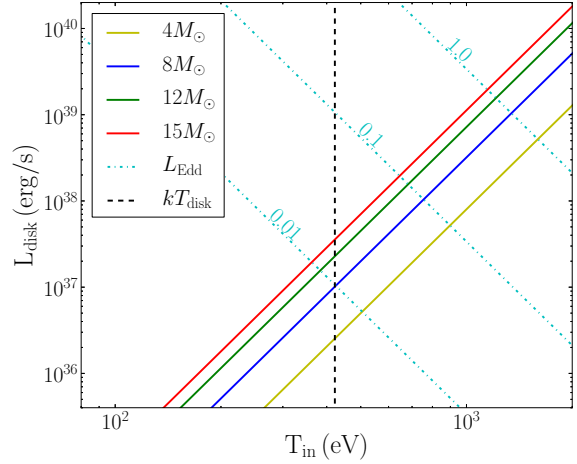}
\caption{Disk luminosity versus disk temperature, allowing an estimate of the source distance. The solid yellow, blue, green and red lines are the expected disk luminosities at each temperature for a range of black hole masses. The dashed cyan lines represent fractions of the Eddington luminosity (where $L_{\rm Edd} = 1.3 \times 10^{38} M/M_{\odot}$) and the black line is our best fitting disk temperature during the softest state.}
\label{fig:source_distance}
\end{figure}

For any given mass, $T_{\rm in}$ corresponds to a specific $L_{\rm disk}$. We select a range of representative masses (4--15 M$_\odot$) from \citet{2012ApJ...757...36K}, typical of Galactic BHs. Based on our measured value of $T_{\rm in}$, that range of masses corresponds to a range of $L_{\rm disk}$. Comparing the predicted range of $L_{\rm disk}$ with our measured flux allows us to estimate the source distance (see Figure~\ref{fig:source_distance}). The estimated source distance for a standard LMXB system (with $T_{\rm in} \approx 0.422$) is between 2.75 kpc (calculated for a 4 $M_{\odot}$ BH) and 10 kpc (for a 15 $M_{\odot}$ BH). We may rule out the lowest distance value as the transition from the hard state to the HIMS is expected at $\ge$ 0.03 $L_{\rm Edd}$ and the transition back occurs between 0.5\% and 10\% $L_{\rm Edd}$ \citep{2010MNRAS.403...61D} with a mean value of $\sim$2\% \citep{2003A&A...409..697M}. We can therefore safely assume that the bolometric disk luminosity at the softest point of the outburst is $> 1\% L_{\rm Edd}$, placing a conservative lower limit of $\sim$~4 kpc to the source.

\section{Summary}
\label{sec:summary}
We have modelled the full radio to X-ray spectra of \source{} during its 2011 outburst. Our simultaneous multiwavelength observations provide an unprecedented insight into the evolution of the accretion flow and corresponding changes within the jet that occur during outburst. DMR13 presented the evolution of the compact jet during the outburst. Here, we have attempted to couple that evolution with the changing accretion flow and compare these results with expected relations. We find that:
\begin{itemize}[itemsep=5pt,topsep=5pt, leftmargin=0.2cm]

\item The radius of the first acceleration zone in the compact jet changes by $\sim$3 orders of magnitude during our observations (from $\sim 10^6 r_{\rm g}$ during the HIMS to $10^3 r_{\rm g}$ early in the outburst decay), suggesting that the jet properties evolve significantly over the course of the outburst. 

\item The evolution of $\nu_{\rm b}$ does not appear to scale with the source luminosity in the hard state, but possibly with hardness ($L_{\rm C}/L_{\rm d}$). However, this is more likely set by much more complex factors and the predicted $\nu_{\rm b} \propto L_{\rm X}^{1/3}$ relation may hold once the system has settled into the established hard state

\item Early in the outburst the high energy synchrotron cooling break occurs in the optical band (between $\approx 3.2 \times 10^{14}$ Hz and $\approx 4.5 \times 10^{14}$ Hz). We speculate that the cooling break shifts to lower frequencies as $\nu_{\rm b}$ also moves to lower frequencies. 

\item The jet accounts for $\sim 6 \%$ of the total energy output of the system early in the outburst, indicating that the system was not jet-dominated at this time. The system may become jet-dominated as the outburst fades if the X-ray power-law emission was from the optically thin synchrotron emission from the jet.

\item The high-energy (hard) X-ray emission may be dominated by the jet during the hard state, as we find similar spectral slopes and normalisation between the optically thin synchrotron emission and the hard X-rays during the decay phase of the outburst.

\item The source distance is 4--10 kpc, which is estimated from the inner-disk temperature, the X-ray state transition luminosity and plausible values of BH mass (4--15 M$_\odot$).
\end{itemize}

Our findings demonstrate the importance of high-cadence multiwavelength observations to better understand the accretion-ejection coupling. In particular, observations at sub-mm and IR wavelengths are crucial to precisely determine the frequency of the jet break and further understand the evolution of the jet during an outburst.

\section*{Acknowledgments}
We would like to thank the anonymous referee for their helpful comments and suggestions. We also thank Christian Motch and Manfred Pakull for their contribution and help with the preparation and execution of the VLT observations, as well as Guillaume Dubus for useful discussions on synchrotron theory. This research was supported under the Australian Research Council's Discovery Projects funding scheme (project number DP 120102393). DMR acknowledges support from a Marie Curie Intra European Fellowship within the 7th European Community Framework Programme under contract no. IEF 274805. The National Radio Astronomy Observatory is a facility of the National Science Foundation operated under cooperative agreement by Associated Universities, Inc. The Submillimeter Array is a joint project between the Smithsonian Astrophysical Observatory and the Academia Sinica Institute of Astronomy and Astrophysics and is funded by the Smithsonian Institution and the Academia Sinica. VLT observations were collected at the European Southern Observatory, Chile, under ESO Programme IDs 087.D-0914 and 089.D-0970.

\newpage

\section*{Appendix}
\FloatBarrier
\begin{table}
\caption{Radio flux density of \source{} for each frequency coincident with other multiwavelength data. Quoted 1$\sigma$ uncertainties are statistical only.}
\label{tab:radio_data}
\renewcommand{\arraystretch}{1.3}
\centering
\begin{tabular}{cccccccccccccc}

\hline

Date & MJD & Frequency  & Flux Density \\
     &     &  (GHz)     &  (mJy) \\
\hline

2011 Sep 3 & 55807.12 & 4.6 & 27.0$\pm$0.6 \\
2011 Sep 3 & 55807.12 & 7.9 & 40.2$\pm$0.9 \\

2011 Sep 12 & 55816.97 & 5.0 & 31.2$\pm$0.7 \\
2011 Sep 12 & 55816.97 & 7.4 & 33.1$\pm$0.7 \\
2011 Sep 12 & 55816.97 & 20.8 & 41.3$\pm$1.5 \\
2011 Sep 12 & 55816.97 & 25.9 & 42.8$\pm$1.5 \\
2011 Sep 12 & 55816.97 & 32.0 & 45.8$\pm$2.5 \\
2011 Sep 12 & 55816.97 & 41.0 & 45.6$\pm$2.8 \\

2011 Sep 17 & 55821.97 & 5.0  & 34.5$\pm$0.9 \\
2011 Sep 17 & 55821.97 & 7.4  & 36.4$\pm$0.9 \\
2011 Sep 17 & 55821.97 & 20.8  & 43.7$\pm$1.7 \\
2011 Sep 17 & 55821.97 & 25.9  & 45.9$\pm$1.8 \\
2011 Sep 17 & 55821.97 & 32.0  & 50.5$\pm$3.0 \\
2011 Sep 17 & 55821.97 & 41.0  & 44.0$\pm$3.2 \\

2011 Sep 26 & 55830.95 & 5.0  & 14.1$\pm$0.4 \\
2011 Sep 26 & 55830.95 & 7.4 & 14.8$\pm$0.4 \\
2011 Sep 26 & 55830.95 & 20.8 & 26.8$\pm$1.1 \\
2011 Sep 26 & 55830.95 & 25.9 & 29.9$\pm$1.2 \\
2011 Sep 26 & 55830.95 & 31.5 & 35.5$\pm$2.1 \\
2011 Sep 26 & 55830.95 & 37.5 & 40.0$\pm$2.5\\

2011 Oct 12 & 55846.01 & 5.3  & 5.5$\pm$0.3 \\
2011 Oct 12 & 55846.01 & 7.4 & 5.9$\pm$0.3 \\
2011 Oct 12 & 55846.01 & 20.8 & 9.8$\pm$0.5 \\
2011 Oct 12 & 55846.01 & 25.9 & 12.2$\pm$0.7 \\
2011 Oct 12 & 55846.01 & 31.5 & 13.5$\pm$1.1\\
2011 Oct 12 & 55846.01 & 37.5 & 13.8$\pm$1.2 \\

2011 Oct 22 & 55856.84 & 5.3 & 3.9$\pm$0.2 \\
2011 Oct 22 & 55856.84 & 7.4 & 4.3$\pm$0.2 \\
2011 Oct 22 & 55856.84 & 20.8 & 5.3$\pm$0.4 \\
2011 Oct 22 & 55856.84 & 25.9 & 5.6$\pm$0.5 \\
2011 Oct 22 & 55856.84 & 31.5 & 5.8$\pm$0.6\\
2011 Oct 22 & 55856.84 & 37.5 & 5.7$\pm$0.8 \\

2011 Nov 1 & 55866.91 & 5.3 & 1.3$\pm$0.2  \\
2011 Nov 1 & 55866.91 & 7.4 & 1.6$\pm$0.2 \\
2011 Nov 1 & 55866.91 & 20.8 & 2.3$\pm$0.4 \\
2011 Nov 1 & 55866.91 & 25.9 & 2.6$\pm$0.5 \\
2011 Nov 1 & 55866.91 & 31.5 & 2.6$\pm$0.5 \\
2011 Nov 1 & 55866.91 & 37.5 & 3.2$\pm$0.7 \\

\end{tabular}
\end{table}

\label{lastpage}

\end{document}